\documentclass[
  aip,
  amsmath,amssymb,
  reprint,
]{revtex4-1}

\usepackage{graphicx}
\usepackage{dcolumn}% Align table columns on decimal point
\usepackage{bm}% bold math

\usepackage[utf8]{inputenc}
\usepackage[T1]{fontenc}
\usepackage{amsmath, mathtools, amssymb}

\usepackage{etoolbox}

\usepackage{mathrsfs}
\usepackage{hyperref} 
\usepackage{makecell}

\makeatletter
\def\@email#1#2{%
  \endgroup
  \patchcmd{\titleblock@produce}
  {\frontmatter@RRAPformat}
  {\frontmatter@RRAPformat{\produce@RRAP{*#1\href{mailto:#2}{#2}}}\frontmatter@RRAPformat}
}

\usepackage{xcolor}
\definecolor{pink}{rgb}{0.8, 0.0, 0.6}

\newcommand{\figeventcamera}{
    \begin{figure}[b]
    \includegraphics[width=0.75\linewidth]{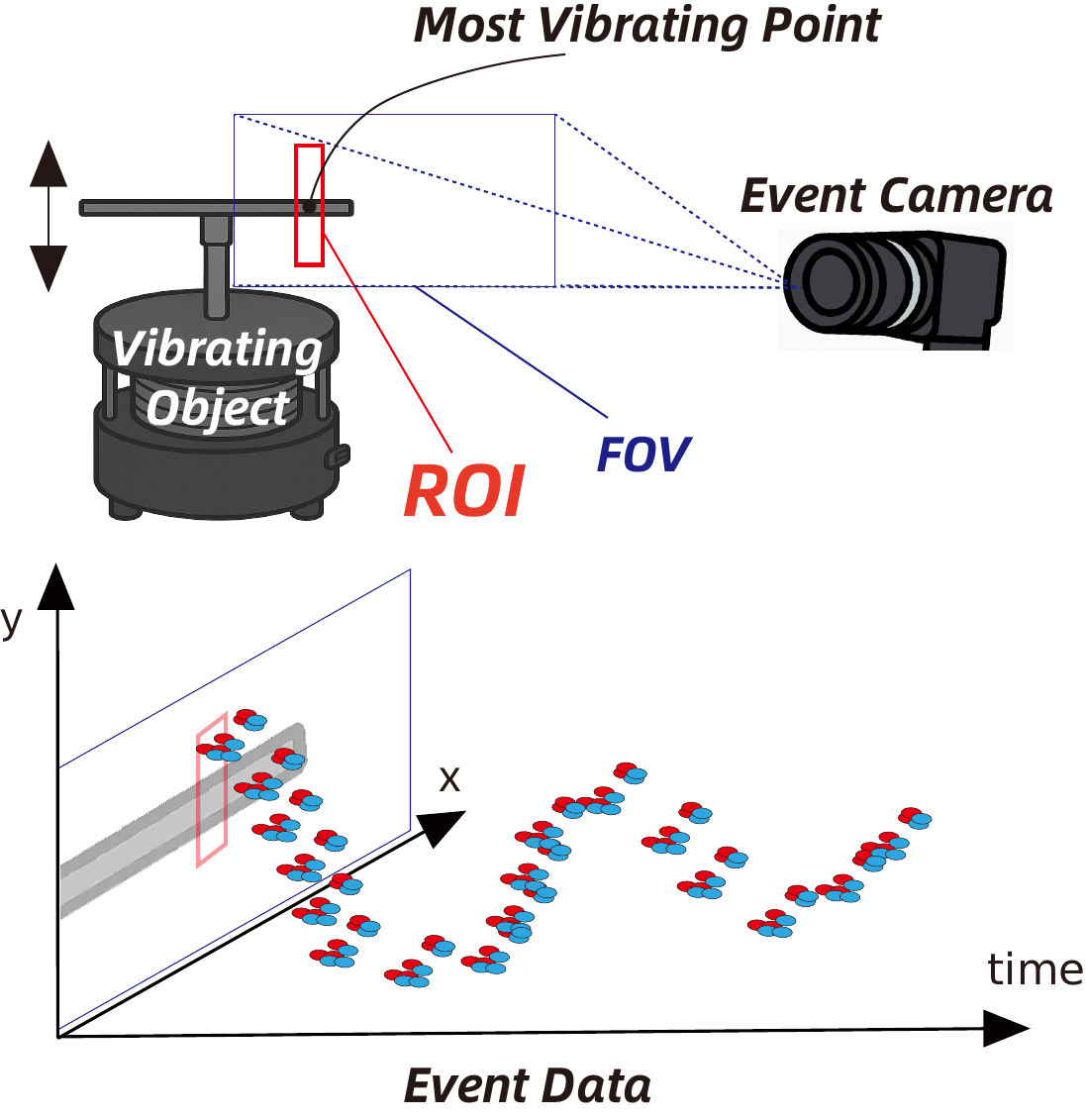}
    \caption{Event camera that records changes in brightness only (red: positive events for brightness increases, blue: negative events for brightness decreases). This function provides a high temporal resolution.
    }
    \label{fig:phasePlate}
    \end{figure}
}

\newcommand{\figoverviewmapper}{
    \begin{figure*}[t]
        \includegraphics[width=\textwidth]{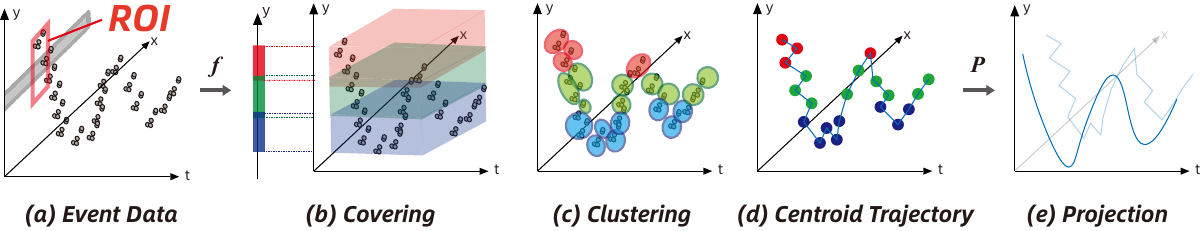}
        \caption{Overview of the proposed vibration reconstruction pipeline. The process begins with (a) recorded data events for target vibration, where raw events are captured as a point cloud in $x$, $y$, and $t$ space. The red box in (a) indicates the region of interest (ROI) used for vibration reconstruction. In our method, the ROI is centered at the pixel with the highest event density and oriented along the vibration direction.
        This event data are then (b) filtered along the $y$-axis and split into multiple intervals, creating distinct temporal segments. Subsequently, (c) HDBSCAN clustering is applied within each interval. 
        In (d), the centroid of each cluster is computed, and these points are chronologically connected by the blue line to form a 3D vibration trajectory. Finally, (e) this trajectory is projected onto the y-t plane, discarding horizontal information to produce the final vibration waveform.}
        \label{fig:overview_of_mapper_algorithm}
    \end{figure*}
}

\newcommand{\figroianalysis}{
    \begin{figure*}[t]
        \centering
        \includegraphics[width=0.8\linewidth]{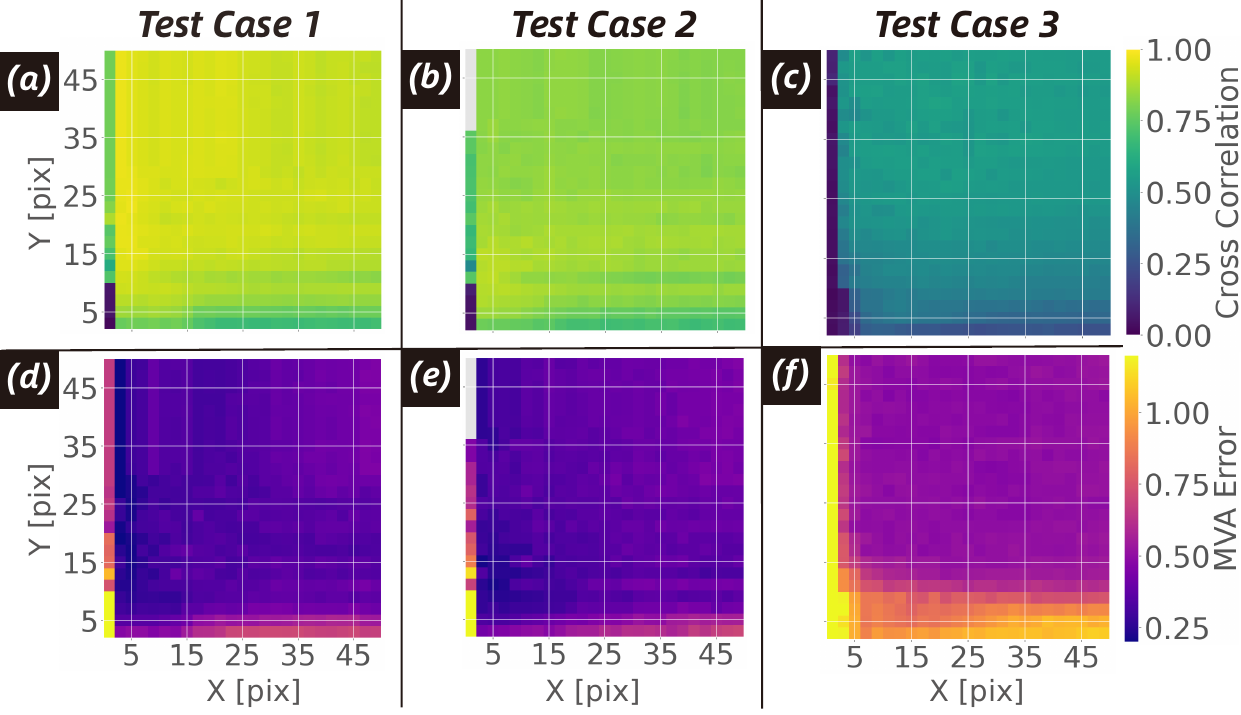}
        \caption{
        Reconstruction accuracy as a function of ROI dimensions (width, height) for three vibration test conditions: Test Case 1 (increasing amplitude signal), Test Case 2 (decreasing amplitude signal), and Test Case 3 (audio signal). (a-c) Normalized cross-correlation maps and (d-f) MVA error maps for Test Cases 1, 2, and 3, respectively.
        }
        \label{fig:roi_analysis}
    \end{figure*}
}

\newcommand{\figresult}{
    \begin{figure*}[t]
        \centering
        \includegraphics[width=\textwidth]{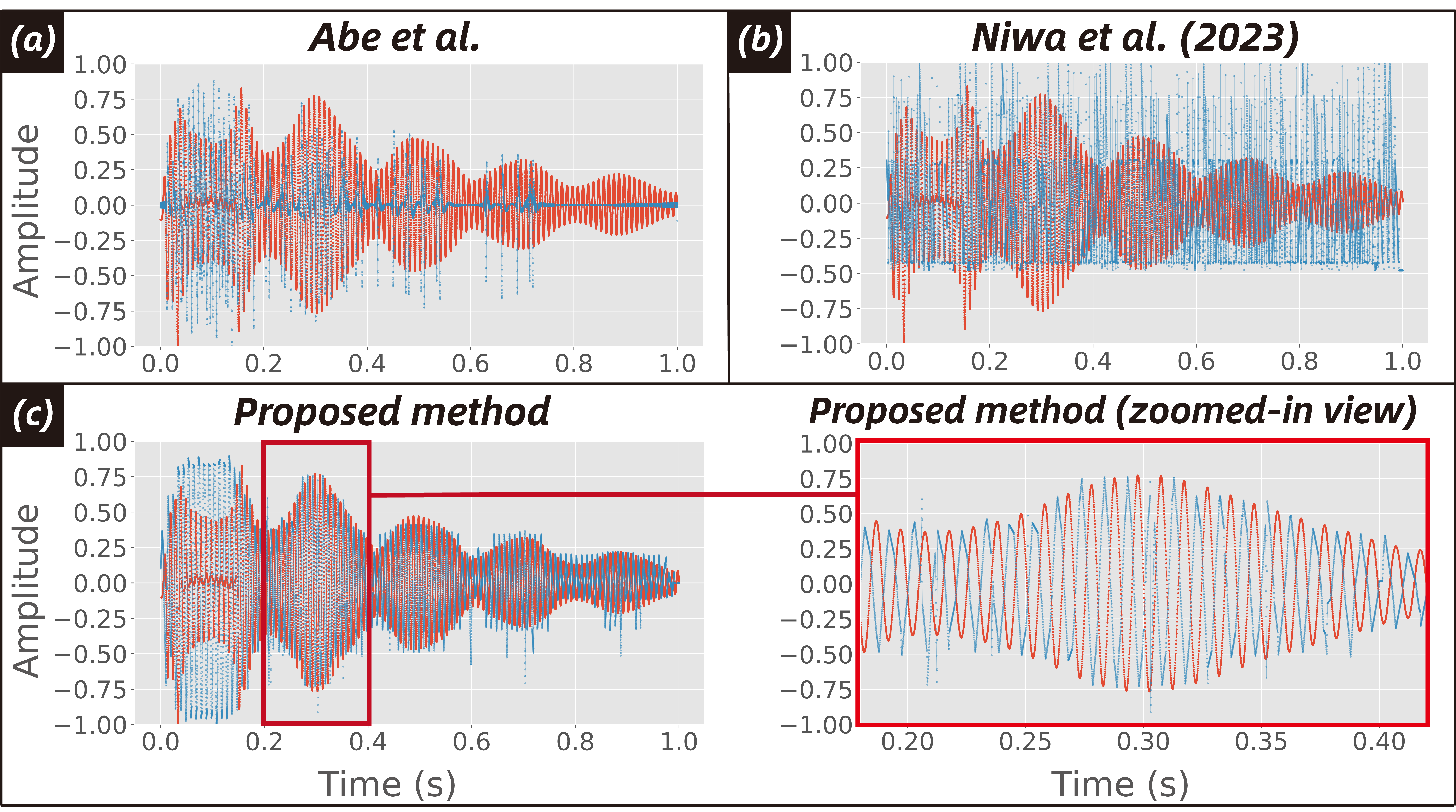}
        \caption{
        Comparison of vibration waveforms reconstructed through various methods and LDV measurements.
        The orange traces in all subplots represent the ground truth vibration measured by an LDV. The blue traces show the reconstructed vibration waveforms from different methods: (a) Abe et al.'s method (frame-based approach), (b) Niwa et al.'s method (event-based phase-based method), and (c) the proposed method. The right panel shows a zoomed-in view of the red box in the left panel.
        }
        \label{fig:result}
    \end{figure*}
}

\newcommand{\figappandresult}{
    \begin{figure}
        \centering
        \includegraphics[width=\linewidth]{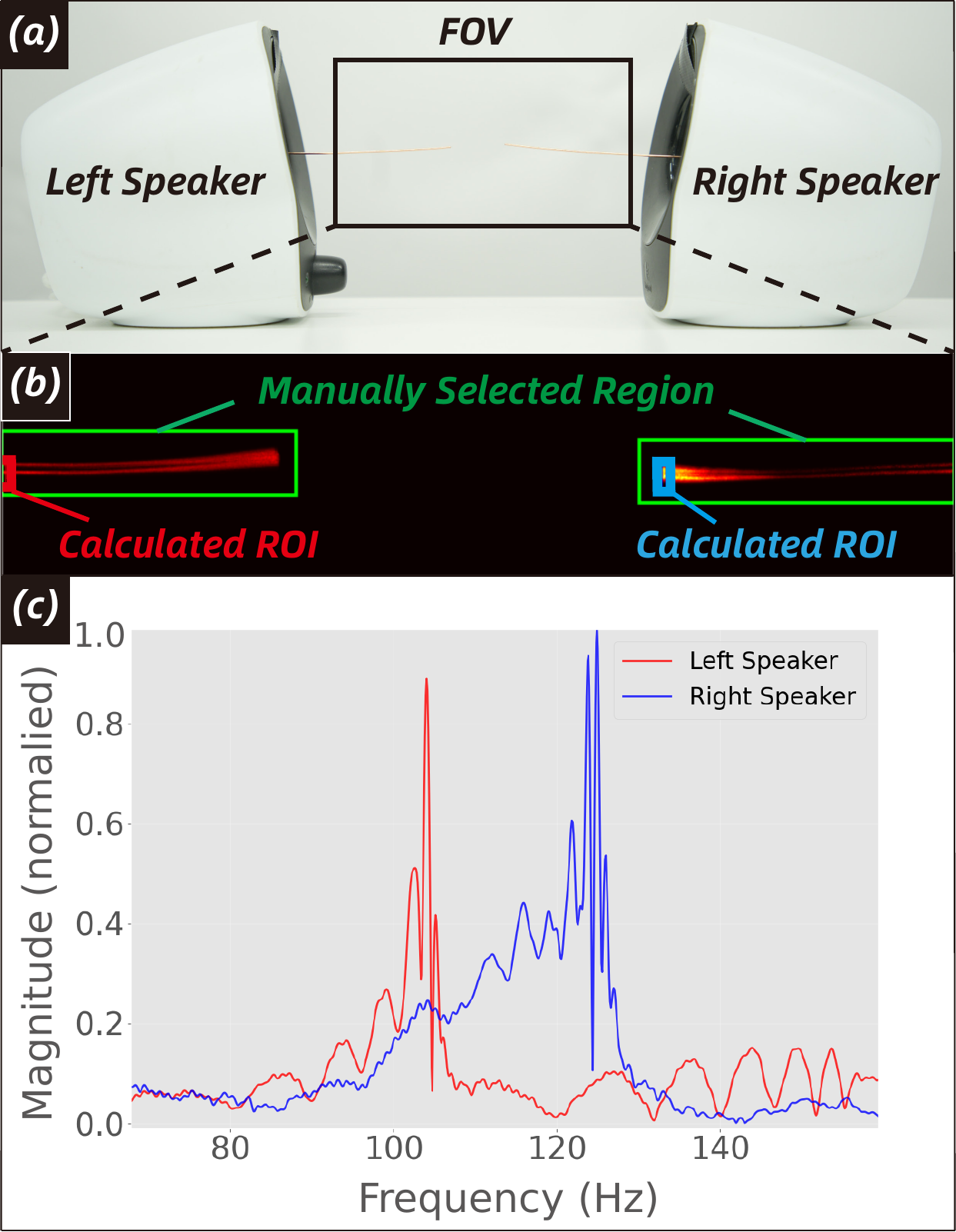}
        \caption{Simultaneous recovery of multiple sound sources. (a) Experimental setup with two speakers playing tone of 100~Hz tone (left) and a 120~Hz tone (right). 
        (b) ROI selection on an accumulated event image. Within the larger, manually selected region (green box), the final calculated ROI (left speaker: red box, right speaker: blue box) is automatically centered on the pixel with the highest event count.
        (c) The FFT of the signals recovered from each speaker's corresponding ROI. The resulting spectra clearly show distinct peaks at 100~Hz (red) and 120~Hz (blue), demonstrating the ability to individually record multiple sources from a single event stream.}
        \label{fig:app_and_result}
    \end{figure}
}

\makeatother
\begin{document}

\preprint{AIP/123-QED}

\title[Event Topology-based Visual Microphone for Amplitude and Frequency Reconstruction]{Event Topology-based Visual Microphone for Amplitude and Frequency Reconstruction}

\author{Ryogo Niwa}
\affiliation{School of Informatics, College of Media Arts, Science and Technology, University of Tsukuba, Kasuga Campus Kasuga 1-2, Tsukuba, Ibaraki, 305-8550, Japan}
\affiliation{R\&D Center for Digital Nature, University of Tsukuba, Tsukuba, 305-8550, Ibaraki, Japan}

\author{Yoichi Ochiai}
\affiliation{R\&D Center for Digital Nature, University of Tsukuba, Tsukuba, 305-8550, Ibaraki, Japan}
\affiliation{Institute of Library, Information and Media Science, University of Tsukuba, Tsukuba, 305-8550, Ibaraki, Japan}
\affiliation{Tsukuba Institute for Advanced Research (TIAR), University of Tsukuba, 1-1-1 Tennodai, Tsukuba, 305-8577, Ibaraki, Japan}
\affiliation{Pixie Dust Technologies, Inc., Chuo-ku, 104-0028, Tokyo, Japan}

\author{Tatsuki Fushimi}
\affiliation{R\&D Center for Digital Nature, University of Tsukuba, Tsukuba, 305-8550, Ibaraki, Japan}
\affiliation{Institute of Library, Information and Media Science, University of Tsukuba, Tsukuba, 305-8550, Ibaraki, Japan}
\affiliation{Tsukuba Institute for Advanced Research (TIAR), University of Tsukuba, 1-1-1 Tennodai, Tsukuba, 305-8577, Ibaraki, Japan}

\email{niwa.ryogo@digitalnature.slis.tsukuba.ac.jp}

\date{\today}

\begin{abstract}
Accurate vibration measurement is vital for analyzing dynamic systems across science and engineering, yet noncontact methods often balance precision against practicality. 
Event cameras offer high-speed, low-light sensing, but existing approaches fail to recover vibration amplitude and frequency with sufficient accuracy.
We present an event topology-based visual microphone that reconstructs vibrations directly from raw event streams without external illumination.
By integrating the Mapper algorithm from topological data analysis with hierarchical density-based clustering, our framework captures the intrinsic structure of event data to recover both amplitude and frequency with high fidelity.
Experiments demonstrate substantial improvements over prior methods and enable simultaneous recovery of multiple sound sources from a single event stream, advancing the frontier of passive, illumination-free vibration sensing.
\end{abstract}

\maketitle

Vibration measurements are critical in fields such as structural health monitoring, product design, industrial diagnostics, and materials science.
Noncontact methods are especially valuable because they avoid mechanical loading on the studied structure.
Laser Doppler vibrometers (LDVs) exhibit high accuracy; however, their cost can limit use outside well-equipped laboratories.
High-speed imaging has been explored as a more accessible alternative \cite{Zalevsky2009-je, Davis2014-oj}.
Nevertheless, operating at very high frame rates often necessitates short exposure times with strong supplemental lighting, which can compromise spatial resolution because of the use of reduced regions of interest or pixel binning.

\figeventcamera

Event cameras are a promising solution for these challenges \cite{TheSiliconRetina, gallego-2022-eventcamera-survey}.
Unlike conventional cameras that capture entire frames at a fixed rate, event cameras operate asynchronously, recording a stream of pixel-level brightness changes with microsecond precision, as discussed in \autoref{fig:phasePlate}.
Responding to relative, rather than absolute, brightness changes allows event cameras to achieve high temporal resolution while maintaining a high spatial resolution without intense illumination, as required by frame-based cameras.
Leveraging these advantages, event cameras have been successfully applied to track fast physical phenomena in related fields such as experimental mechanics \cite{Raynaud2024-dn} and airborne particle control \cite{Ren2022-xy}.

\figoverviewmapper

Event cameras are also particularly well suited for vibration measurement owing to their ability to capture fast dynamics. 
Approaches for vibration reconstruction using an event camera can broadly be categorized as active or passive capture. 
Active methods, which project an external light source onto a target, can realize high-sensitivity recordings.
However, their use of lasers introduces limitations \cite{yin2025evmiceventbasednoncontactsound, cai2025event2audioeventbasedopticalvibration}. 
Although this approach is precise for single-point measurements, its reliance on capturing laser speckle patterns introduces practical limitations.
For instance, the need to defocus the laser makes it difficult to record multiple vibration sources simultaneously, preventing the full utilization of the event camera's high spatial resolution.
Thus, we focus on passive methods, which record the target without using a strong light source.
One approach to vibration reconstruction is interpreting the times of specular reflections in event data as signal level crossings and recovering the signal through short-time Fourier sparsity \cite{Howard2023-km, Howard:2025:TPAMI}.
In our previous work, we also reconstructed vibrations by generating pseudo-frames from the event data and extracting phase information using Gabor filters \cite{Niwa_undated-ke}.
Although these passive methods can successfully reconstruct the principal frequency components of a vibration, they fail to accurately recover vibration amplitudes.

% Method
In contrast to the methods that rely on first converting events into other representations, our event topology-based visual microphone directly reconstructs the vibration trajectory from the geometric structure of the event stream (see \autoref{fig:overview_of_mapper_algorithm}). 
To achieve this, we leverage techniques from Topological Data Analysis (TDA), a field of mathematics designed to extract robust structural features from complex and high-dimensional datasets.
Specifically, our method adapts the core pipeline of the Mapper algorithm, a primary tool in TDA. 
While the standard Mapper algorithm generates a simplified, graph-based representation of data, we leverage its sequence of covering and clustering to directly reconstruct a vibration.
This direct, topology-aware reconstruction allows us to overcome the challenge of recovering vibration amplitude, a key limitation of prior techniques, and enables applications such as the simultaneous analysis of spatially distinct sources.

The proposed reconstruction process operates on events within a defined region of interest (ROI) that encompasses the vibrating target.
After defining the ROI, we process only the positive events to reduce data redundancy. 
Positive events are triggered when pixel brightness increases, while negative events correspond to reductions in brightness and are discarded.
We then reduce noise in the positive event data using the background activity filter with a parameter \( T \) set to 200~$\mu$s \cite{Delbruck2008-ba-filder}. 
The TDA Mapper algorithm is then applied to the resulting noise-reduced event stream to extract the vibration trajectory.

This reconstruction process, based on the Mapper pipeline, has several stages.
First, the spatiotemporal event data $X \subset \mathbb{R}^3$ (see \autoref{fig:overview_of_mapper_algorithm}(a)) is partitioned into several overlapping subsets. 
This is achieved by first mapping the data onto a one-dimensional space via a filter function, $f: X \rightarrow \mathbb{R}$, and then covering the range of this function with a set of overlapping intervals. 
This combined process is visualized in \autoref{fig:overview_of_mapper_algorithm}(b). 
In our implementation, we use the vertical position ($y$-coordinate) as the filter function, $f(t, x, y) = y$, corresponding to the primary axis of vibration. 
Each resulting subset of events is enclosed in a color-coded cuboid in the figure, defining a specific region for local clustering.

Next, a clustering algorithm is applied to the subset of events within each colored interval.
This process partitions the events into distinct clusters.
For this task, we employ HDBSCAN (Hierarchical Density-Based Spatial Clustering of Applications with Noise) \cite{campello-2013-hdbscan}.
Unlike conventional density-based clustering such as DBSCAN \cite{ester1996-dbscan}, which requires a manually tuned distance threshold $\varepsilon$, HDBSCAN adapts to varying event densities.
This adaptability is crucial for robustly handling complex vibrations where frequency and amplitude changes alter the local event density. 
\autoref{fig:overview_of_mapper_algorithm}(c) illustrates the outcome of this stage, showing the identified clusters after HDBSCAN has filtered out sparse events classified as noise.
The interval and HDBSCAN parameters are detailed in the supplementary material.

\figroianalysis

At this stage, our approach departs from the standard Mapper algorithm. 
Instead of constructing a graph-based representation to capture the data's topology, we compute the centroid of each cluster to form a discrete representation of the vibration trajectory.
Subsequently, each cluster identified in \autoref{fig:overview_of_mapper_algorithm}(c) is represented by its centroid, computed as the mean of the coordinates of all events within the cluster in the original $\mathbb{R}^3$ space.
By temporally connecting these centroids, we reconstruct a three-dimensional spatiotemporal trajectory, depicted as the line connecting the centroids in \autoref{fig:overview_of_mapper_algorithm}(d).
Although this 3D trajectory captures the in-plane motion, our primary interest is in the vertical vibration.
Therefore, we apply a projection $P: \mathbb{R}^3 \rightarrow \mathbb{R}^2$ defined by $P(t, x, y) = (t, y)$, intentionally discarding the horizontal information.
This projection yields a fundamental representation of vibration in physics: a waveform plotting displacement ($y$) as a function of time ($t$), as shown in \autoref{fig:overview_of_mapper_algorithm}(e).
This one-dimensional signal is then resampled to a uniform rate of 16~kHz using linear interpolation.

To evaluate the performance of our system, we conducted experiments using an aluminum plate ($10~\text{mm} \times 100~\text{mm}$, $0.3~\text{mm}$ thick) that was vertically actuated by a mechanical wave driver (WA-9855) via an applied voltage (see \autoref{fig:phasePlate}), under three conditions: (1) Test Case 1, a 100~Hz sinusoid with increasing amplitude; (2) Test Case 2, a 100~Hz sinusoid with decreasing amplitude; and (3) Test Case 3, a human speech signal reciting \textit{``Mary had a little lamb.''}
The vibrations were recorded with an event camera (SilkyEvCam HD, EVK3) equipped with an 8~mm lens at a distance of 20~cm from the target, and simultaneously with an LDV (Polytec-500-3D-HV-Xtra) to obtain ground-truth measurements. 
For this setup, the imaging scale at the object plane was approximately 0.12~mm/pixel.
See the supplementary material for more details on the experimental setup.

\figresult

To evaluate the reconstruction accuracy, we employ two primary metrics.
Because the reconstructed signal and the ground-truth signal possess different physical units, the standard signal-to-noise ratio (SNR) is not an appropriate measure of similarity.
Instead, we first compute the normalized cross-correlation to assess waveform fidelity. 
This metric offers a robust measure of similarity that remains invariant to differences in signal scale and DC offset.
To further assess spectral fidelity, we compute the Median Vector Angle (MVA) error, following the methodology of Howard et al. \cite{Howard2023-km, Howard:2025:TPAMI}.
This metric quantifies spectral similarity while remaining robust to differences in the scale of the amplitude and the global phase.
A lower MVA error indicates a higher correlation in the frequency domain.
The calculation first determines the cosine similarity, $M(j)$, between the short-time Fourier transform of the reconstructed signal, $\mathcal{G}(j, \omega)$, and that of the ground-truth reference signal, $\mathcal{F}(j, \omega)$, for each short-time index $j$.
\begin{equation}
    M(j)=\frac{\int|\mathcal{G}(j, \omega) \mathcal{F}(j, \omega)| d \omega}{\sqrt{\int|\mathcal{G}(j, \omega)|^2 d \omega} \sqrt{\int|\mathcal{F}(j, \omega)|^2 d \omega}}
\end{equation}
The corresponding vector angle error is then computed by applying the arccosine to $M(j)$:
\begin{equation}
    A(j)=\cos^{-1}(M(j))
\end{equation}
The reported MVA error corresponds to the median of the $A(j)$ values computed across all time indices.

A key factor influencing reconstruction fidelity is the selection of the ROI.
An ROI that is too small may fail to capture enough events to form stable clusters, while an excessively large ROI can introduce noise and degrade the quality of reconstruction.
To resolve this trade-off and characterize the robustness of our method, we first investigated how reconstruction accuracy varies with ROI size, as shown in \autoref{fig:roi_analysis}.
For this analysis, the ROI center was fixed at the pixel with the highest event density (i.e., the most strongly vibrating point).
The ROI dimensions were then expanded symmetrically from this center. 
For example, a width of 1 includes only the central pixel, whereas a width of 3 encompasses the central pixel plus its immediate horizontal neighbors ($\pm$1 pixel). 
The analysis of both normalized cross-correlation and MVA error revealed a consistent trend across all signal types. Performance degraded noticeably for very narrow ROIs, particularly with a width of 1 pixel, indicating insufficient spatial information.
This degradation is thought be due to an insufficient number of events in this narrow area, preventing the HDBSCAN algorithm from reliably forming clusters. 
This led to reconstruction failures, which manifested as data gaps in \autoref{fig:roi_analysis}(b) and (e) (e.g., at $x = 1$, $y = 37$--$49$~pixels).
In contrast, ROI widths of 3 pixels or more, performance stabilized and became largely insensitive to further increases in width, provided that the ROI height was sufficient. 
A height of approximately 25 pixels proved adequate to ensure robust performance.
Based on these observations, we selected a representative ROI of (5, 25) pixels (width, height) for subsequent comparisons. 
This configuration strikes a practical balance between reconstruction fidelity and noise suppression while keeping computational demands manageable.

Using the optimized ROI, we evaluated our method against two baselines: a passive event-based approach (Niwa et al. \cite{Niwa_undated-ke}) and a frame-based method (Davis et al. \cite{Davis2014-oj}), both of which are foundational works in event- and frame-based vibration sensing.
Our focus was on measuring vibrations on diffuse-reflective surfaces, which are common in real-world scenarios. 
Because our target exhibits diffuse rather than specular reflection, the method of Howard et al. \cite{Howard2023-km, Howard:2025:TPAMI}, which relies on specular reflection, was not applicable and was therefore excluded from comparison.
For the frame-based baseline (Davis et al.), we converted the event data into images by generating accumulated event frames, which is a standard technique in which events are integrated over time \cite{gallego-2022-eventcamera-survey}.
To facilitate direct visual comparison in \autoref{fig:result}, all reconstructed waveforms were mean-centered and normalized to the range $[-1, 1]$.
This preprocessing step that does not affect the quantitative evaluation, as both normalized cross-correlation and MVA error are invariant to such linear transformations.

Our proposed method outperforms the baseline techniques. 
This improvement is qualitatively visible in \autoref{fig:result}: under one vibration condition, our method accurately reconstructs the envelope of the reference waveform measured by the LDV, in contrast to the baseline approaches. 
Results for the other two vibration types are provided in the supplementary material.
These qualitative observations are corroborated by quantitative results in \autoref{table:compare_by_xcorr} and \autoref{table:compare_by_MVA}. 
Across all conditions, our method achieves the highest normalized cross-correlation scores, reflecting more accurate reconstruction of waveform shape.
Furthermore, it consistently yields the lowest MVA error across all three conditions, with values ranging from 0.22--0.53~rad.

\begin{table}[h]
\centering
\caption{Comparison of methods by normalized cross-correlation}
\label{table:compare_by_xcorr}
\begin{tabular}{c|c|c|c}
\hline
Signal & Abe et al. & \makecell{Niwa et al. \\ (2023)} & \textbf{Ours} \\
\hline
Increasing & 0.23 & 0.23 & \textbf{0.97} \\
\hline
Decreasing & 0.34 & 0.29 & \textbf{0.85} \\
\hline
Voice & 0.08 & 0.08 & \textbf{0.46} \\
\hline
\end{tabular}
\end{table}

\begin{table}[h]
\centering
\caption{Comparison of methods by MVA error (rad)}
\label{table:compare_by_MVA}
\begin{tabular}{c|c|c|c}
\hline
Method & Abe et al. & \makecell{Niwa et al. \\ (2023)} & \textbf{Ours} \\
\hline
Increasing & 1.32 & 1.18 & \textbf{0.22} \\
\hline
Decreasing & 1.34 & 1.13 & \textbf{0.33} \\
\hline
Audio & 1.54 & 1.14 & \textbf{0.53} \\
\hline
\end{tabular}
\end{table}

Adapting the frame-based algorithm of Davis \textit{et al.} \cite{Davis2014-oj} to event data using accumulated event frames proved unsuccessful due to a fundamental physical mismatch. 
Their method assumes that pixel values correspond to surface luminance, whereas in accumulated event frames the values instead represent motion intensity (i.e., event counts).
This inconsistency in the underlying physical model prevented accurate reconstruction of both vibration amplitude and frequency, underscoring the need for a dedicated event-based approach.
The method of Niwa \textit{et al.} \cite{Niwa_undated-ke}, which interprets the Gabor-filter phase as a direct proxy for vibration amplitude, is similarly constrained by a fundamental limitation: the filter phase does not preserve a one-to-one correspondence with physical displacement. 
As a result, the method suffers from significant performance degradation, with MVA errors two to five times higher than those achieved by our approach.

It should be emphasized that although the experiments in this study focused on vibrations along a single orientation, the proposed method is not inherently limited in this respect. 
In principle, vibrations occurring along an arbitrary axis can be accommodated by applying a standard rotational transformation as a preprocessing step, aligning the principal axis of motion with the vertical ($y$) axis.
A more fundamental limitation arises in the case of out-of-plane motion---that is, vibrations along the optical axis.
Such motion produces little to no displacement in the 2D image projection, which is the essential signal required for our TDA-based trajectory reconstruction. 
Addressing this limitation is a promising direction for future work. 
One potential avenue is the incorporation of stereo event-based imaging, which could enable the effective measurement of out-of-plane components.

\figappandresult

% application
Finally, we investigated a practical application of our system: the ability to simultaneously recover sound from multiple, spatially distinct sources using a single event stream. 
As illustrated in \autoref{fig:app_and_result}(a), we conducted an experiment with two speakers, playing a 100~Hz tone from the left speaker and a 120~Hz tone from the right speaker simultaneously. 
To provide distinct visual targets, a $0.8$~mm diameter metal rod was attached to the cone of each speaker.
For each rod, the final $5 \times 25$~pixels ROI was determined using our centering procedure, applied within a larger manually selected region (see accumulated event image in \autoref{fig:app_and_result}(b)).
The FFTs of the reconstructed signals are shown in \autoref{fig:app_and_result}(c), with the red and blue lines corresponding to the left and right speakers, respectively.
Although the detected peaks exhibit slight frequency shifts from the nominal 100~Hz and 120~Hz, two clearly separated peaks are observed, corresponding to the individual sources.
These results demonstrate that each reconstruction faithfully captures only its respective frequency component. 
This capability enables our method to record multiple sound sources independently---something a single conventional microphone cannot achieve, as it would instead capture only a mixed signal.

% conclusion
We presented an event topology-based visual microphone, a passive event-camera-based approach capable of reconstructing both frequency and amplitude.
By integrating a Mapper-based framework with HDBSCAN, our approach achieved stable cluster identification, allowing faithful reconstruction even at the various event densities associated with complex or nonstationary vibrations.
Across all tested signals, our method consistently outperformed existing baselines, achieving the highest normalized cross-correlation and the lowest MVA error.
A fundamental limitation is the inability to measure out-of-plane motion along the optical axis because such vibrations do not generate a significant displacement in the 2D image plane.
Future work will focus on addressing this limitation, for example by incorporating stereo event-based imaging, and on validating the method in a broader range of real-world applications.

\begin{acknowledgments}
This work was supported by Grant-in-Aid for Scientific Research (No. 24KJ0497).
\end{acknowledgments}

\bibliography{references}

\end{document}